\documentclass[sigconf]{acmart}

\AtBeginDocument{%
\providecommand\BibTeX{{%
		\normalfont B\kern-0.5em{\scshape i\kern-0.25em b}\kern-0.8em\TeX}}}

\setcopyright{acmcopyright}
\copyrightyear{2021}
\acmYear{2021}
\acmDOI{10.1145/1122445.1122456}

\acmConference[Sigir '21]{The 44th International ACM SIGIR Conference
on Research and Development in Information Retrieval}{July 11--15, 2021}{Online}
\acmBooktitle{The 44th International ACM SIGIR Conference on Research and Development in Information Retrieval,
July 11--15, 2021, Online}
\acmPrice{15.00}
\acmISBN{978-1-4503-XXXX-X/18/06}

\usepackage{multirow}
\usepackage{color}
\usepackage{caption}
\usepackage{subcaption}
\usepackage{float}
\usepackage{enumitem}
\usepackage{xspace}
\usepackage{url}

\newcommand{\modelname}{\textsf{ConsisRec}\xspace}

\setlist[itemize]{leftmargin=*}

\copyrightyear{2021}
\acmYear{2021}
\setcopyright{acmlicensed}\acmConference[SIGIR '21]{Proceedings of the 44th International ACM SIGIR Conference on Research and Development in Information Retrieval}{July 11--15, 2021}{Virtual Event, Canada}
\acmBooktitle{Proceedings of the 44th International ACM SIGIR Conference on Research and Development in Information Retrieval (SIGIR '21), July 11--15, 2021, Virtual Event, Canada}
\acmPrice{15.00}
\acmDOI{10.1145/3404835.3463028}
\acmISBN{978-1-4503-8037-9/21/07}

\begin{document}
\fancyhead{}
\pagenumbering{gobble}

\title{ConsisRec: Enhancing GNN for Social Recommendation via Consistent Neighbor Aggregation}


\author{Liangwei Yang, Zhiwei Liu, Yingtong Dou}
\affiliation{
\institution{Department of Computer Science,}
\city{University of Illinois at Chicago}}
\email{{lyang84, zliu213, ydou5}@uic.edu}

\author{Jing Ma}
\affiliation{
\institution{Business School,}
\city{Sichuan University}}
\email{jingma@stu.scu.edu.cn}

\author{Philip S. Yu}
\affiliation{
\institution{Department of Computer Science,}
\city{University of Illinois at Chicago}}
\email{psyu@uic.edu}



\begin{abstract}
	Social recommendation aims to fuse social links with user-item interactions to alleviate the cold-start problem for rating prediction. Recent developments of Graph Neural Networks~(GNNs) motivate endeavors to design GNN-based social recommendation frameworks to aggregate both social and user-item interaction information simultaneously. However, most existing methods neglect the social inconsistency problem, which intuitively suggests that social links are not necessarily consistent with the rating prediction process. Social inconsistency can be observed from both context-level and relation-level. Therefore, we intend to empower the GNN model with the ability to tackle the social inconsistency problem. We propose to sample consistent neighbors by relating sampling probability with consistency scores between neighbors. Besides, we employ the relation attention mechanism to assign consistent relations with high importance factors for aggregation. Experiments on two real-world datasets verify the model effectiveness.
	
\end{abstract}




\maketitle

\section{Introduction}

A recommender system predicts how likely a user is interested in an item \cite{sun2013recommendations,sun2018attentive,liu2019jscn,liu2020basconv,liu2020basket}. However, due to the high cost of data collection, most existing recommender systems suffer from the cold-start problem~\cite{liu2021augmenting}. To alleviate it, we can incorporate the social information among users~\cite{li2020efficient,wu2020diffnet++,wu2018collaborative}, which functions as a side information of the user-item interaction. Previous endeavour~\cite{bizzi2019double} shows that users' online behaviors are greatly influenced by their social networks, such as the friendship on Wechat~\cite{wu2019dual}, following links on Twitter~\cite{lin2013addressing} and trusting links on shopping website~\cite{fan2020graph}. 
Therefore, fusing the social links with user-item interactions is advantageous to improve the recommendation performance, which is defined as the social recommendation problem.

The recent developments of Graph Neural Networks (GNNs)~\cite{liu2020alleviating,liu2020basket} help handle social recommendation tasks by simultaneously aggregating the information from both social graph and user-item graph~\cite{guo2020deep,mu2019graph}. Based on the assumption that neighbors share similar contexts, GNN learns node embeddings by aggregating neighbor information recursively on graphs~\cite{dou2020enhancing}. SocialGCN~\cite{wu2018socialgcn,wu2019neural} proposes to enhance user embedding by simulating how users are influenced by the recursive social diffusion process. GraphRec~\cite{fan2019graph} and GraphRec+~\cite{fan2020graph} jointly model three types of aggregation upon social graph, user-item graph and item-item graph, to learn user\&item embeddings comprehensively. DSCF~\cite{fan2019deep} includes high-order social links through a sequential learning on random walks. MGNN~\cite{xiao2020mgnn} builds mutual social embedding layers to aggregate information from user-item rating graph and social graph. 

\begin{figure*}[!hbt]
    \centering
    \includegraphics[width=0.8\linewidth]{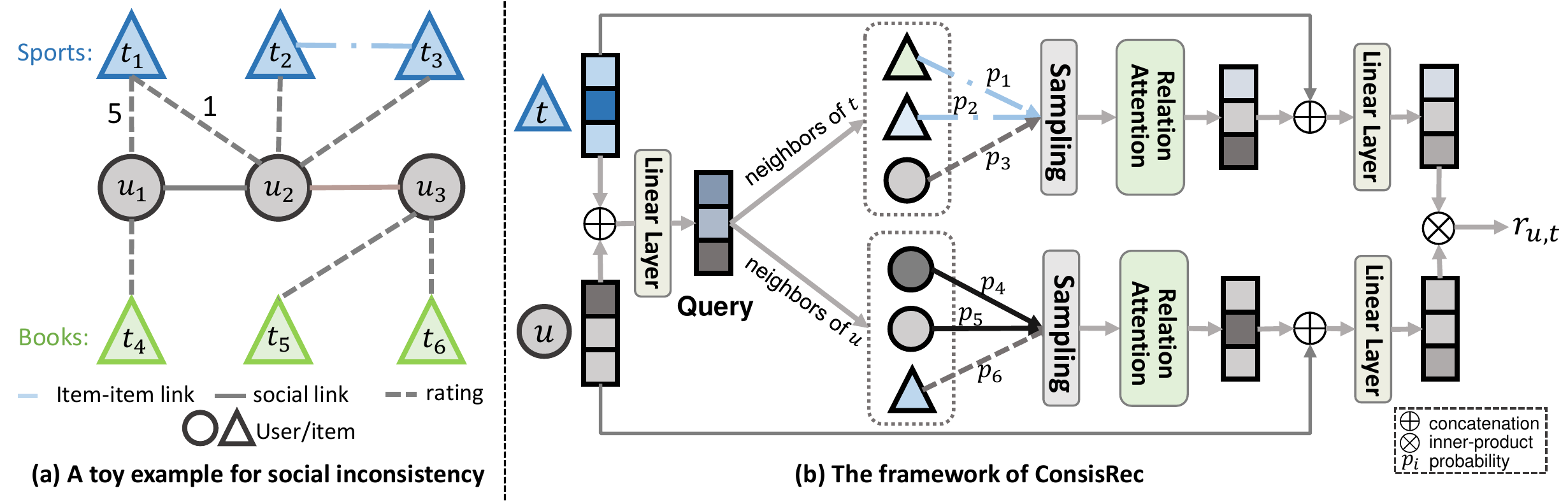}
    \caption{A toy example reflecting social inconsistency and the framework of \modelname.}
    \label{framework}
\end{figure*}

However, most existing GNN-based social recommendation models ignore the \textbf{social inconsistency} problem~\cite{liu2020alleviating}.
Specifically, the social inconsistency suggests that social links are not necessarily consistent with the rating prediction process. Aggregating the information from inconsistent social neighbors spoils the ability of a GNN to characterize beneficial information for recommendation.
The social inconsistency can be categorized into two levels:
\begin{itemize}
    \item \textit{Context-level:} It indicates that users connected in a social graph may have discrepant item contexts. We demonstrate the context-level social inconsistency in Figure~\ref{framework}(a). We use dash lines and the solid lines to represent user-item ratings and social connections, respectively. As seen, $u_3$ would be $u_2$'s inconsistent neighbor because the items of $u_3$ are all \textit{books}, while $u_2$'s rated items all belongs to \textit{sports}. They have rather discrepant item contexts.
    \item \textit{Relation-level:} There are multiple relations when simultaneously modeling social graph and user-item graph. For example, besides social relations, we also distinguish user-item relations by their rating values. In Figure~\ref{framework}(a), we observe the $u_1$ and $u_2$ are social neighbors and both connected with $t_1$. However, $u_1$ highly likes $t_1$ ($5$ score) while $u_2$ dislikes it ($1$ score). It leads to the relation-level inconsistency because though socially connected, they are of inconsistent item preference.
\end{itemize}

To this end, we intend to empower the GNN model to solve the social inconsistency problem, which is non-trivial. On the one hand, the contexts for both users and items are rather complex and difficult to express explicitly.
On the other hand, we should model multiple relations simultaneously and distinguish the consistent neighbors.
Therefore, we propose a novel framework to tackle the social inconsistency problem when conducting social recommendation, which is named as \modelname.
It is built upon a GNN model~\cite{kipf2016semi}, aggregating neighbors to learn node embeddings. To alleviate the social inconsistency problem, \modelname first generates a query embedding for selecting consistent neighbors. Then, it employs a neighbor sampling strategy for the selection process, where the sampling probability is based on our proposed consistency scores between the query embedding and neighbor embeddings. After sampling, it adopts relation attention to tackle the relation-level inconsistency. As such, neighbors with consistent relations are assigned with high importance factors for aggregation. Therefore, the learned node embeddings for rating prediction are aggregated from consistent contexts and relations. The code is available online at \textcolor{blue}{\url{https://github.com/YangLiangwei/ConsisRec}}. 
The contributions of this paper are listed as follows:

\begin{itemize}
	\item To the best of our knowledge, we are the first work empowering the GNN model to tackle the social inconsistency problem when conducting social recommendation.
	\item We propose a novel framework, \modelname, to learn consistent node embeddings for rating prediction.
	\item Experiments on two real-world datasets show the effectiveness of \modelname. Detailed analyses of \modelname justify its efficacy.
\end{itemize}

\section{Preliminaries}\label{Problem Definition}

Social recommendation problem consists of two set of entities, a user set $\mathcal{U}=\{u_1,u_2,...,u_m\}$ and an item set $\mathcal{T}=\{t_1,t_2,...,t_n\}$, where $m$ and $n$ are the total number of users and items, respectively. It includes two types of information, an incomplete rating matrix $\mathbf{R} \in \mathbb{R}^{m \times n}$ and the user-user social graph $\mathcal{G}_{s} = \{\mathcal{U},\mathcal{E}_{s}\}$. The rating $\mathbf{R}_{u,t} $ denotes user $u$'s preference to item $t$, where higher score is interpreted as more preferring. An social edge $(u_i,u_j) \in \mathcal{E}_{s}$ indicates that user $i$ has an social connection with $j$, \textit{e.g.}, trust.

The objective of social recommendation is to complete the rating matrix by fusing the rating matrix and the social graph. 
Therefore, we solve social recommendation problem by constructing a heterogeneous graph $\mathcal{G}=\{\mathcal{V}, {\mathcal{E}_r}|_{r=1}^R\}$, where $\mathcal{V}$ denotes both user and item nodes and $\mathcal{E}_{r}$ denotes edges upon relation $r$.  Besides user-user and item-item links, we also distinguish the user-item links by their rating scores. The score set varies on different datasets. \textit{E.g.}, Ciao dataset \cite{tang2012etrust} has 6 rating values, \textit{i.e.,} $\{0,1,2,3,4,5\}$. Hence, the edges on Ciao have $8$ types, \textit{i.e.,} $R=8$, one being social relation, one being item-item relation and the others being different rating values.

\section{Proposed Model}

The framework of \modelname is shown in Figure \ref{framework}(b)\footnote{Although we only illustrate the one-layer version of \modelname in the figure, our model is flexible to have $L$ layers as introduced in the following section.}. It has \textit{embedding layer}, \textit{query layer}, \textit{neighbor sampling} and \textit{relation attention}.

\subsection{Embedding Layer}\label{query embedding}
Following existing works~\cite{wu2019neural,lightgcn2020he}, we maintain an embedding layer $\mathbf{E}\in\mathbb{R}^{d\times (m+n)}$, each column of which represents the trainable embedding for each node.
We can index it to retrieve the embedding of a node $v\in \mathcal{U}\cup\mathcal{T}$ as $\mathbf{e}_{v}\in \mathbb{R}^{d}$. In the following sections, without specific statements, we use an index $v$ to denote a node, which can either be a user or an item, while $u$ and $t$ specifically denote a user and an item node.
Apart from node embeddings, we also train a relation embedding vector for each relation $r$ to characterize \textit{relation-level} social inconsistency, denoted as $\mathbf{e}_{r}$.  

\subsection{Query Layer}
To overcome the social inconsistency problem, we should aggregate consistent neighbors to learn node embeddings. Since social inconsistency are both in context-level and relation-level, we should distinguish consistent neighbors for each pair $(u,t)$.
Therefore, \modelname employs a query layer to exclusively select consistent neighbors for the query pair $(u,t)$. It generates a query embedding by mapping the concatenation of user and item embeddings:
\begin{equation}
	\mathbf{q}_{u,t}=\sigma\left(\mathbf{W}_{q}^{\top}(\mathbf{e}_{u}\oplus \mathbf{e}_{t})\right),
\end{equation}
where $\mathbf{q}_{u,t}$ is the query embedding , $\mathbf{e}_{u},\mathbf{e}_{t}\in \mathbb{R}^{d}$ are the embedding for node $u$ and $t$, respectively, $\oplus$ denotes concatenation, $\mathbf{W}_{q}\in\mathbb{R}^{2d\times d}$ is the mapping matrix, and $\sigma$ is a ReLU activation function. We design a query layer to dynamically sample neighbors based on different items. It is because when users buy different items, they would inquire different friends. Thus, $u$'s rating score of $t$ is related to friends who are familiar with this query item $t$. 


\subsection{Neighbor Sampling}
Neighbor sampling has been applied to GNN to boost training~\cite{cong2020minimal,chiang2019cluster,zeng2019graphsaint} and improve ranking performance~\cite{najork2009less}. Compared with previous work, ConsisRec aims to deal with the inconsistency problem in social recommendation, and dynamic samples different social neighbors based on different items. Next, we present how to sample neighbors for learning the embedding of $u$ and $t$. The framework of \modelname to aggregate node embeddings can be formalized as:
\begin{equation}\label{eq:GNN}
	\mathbf{h}_v^{(l)}= \sigma\left(\mathbf{W}^{(l)\top}\left(\mathbf{h}_v^{(l-1)} \oplus \text{AGG}^{(l)}\{\mathbf{h}_{i}^{(l-1)} | i\in{\mathcal{N}_v}\}\right)\right),
\end{equation}
where $\sigma$ is a ReLU activation function, $\mathbf{h}_v^{(l)}\in\mathbb{R}^{d}$ is the hidden embedding of node $v$ at $l$-th layer, $\mathcal{N}_v$ is the sampled neighbors of node $v$, AGG is an aggregation function, and $\mathbf{W}^{(l)}\in \mathbb{R}^{2d\times d}$ is the mapping function. $\mathbf{h}_v^{(0)}$ is the initial node embedding of $v$, \textit{i.e.,} $\mathbf{e}_v$. 

Instead of equally aggregating all neighbors, we should emphasize more on consistent neighbors while ignoring those inconsistent neighbors. Therefore, we propose to use neighbor sampling method to select those consistent neighbors. The sampling probability for neighbor node $i$ at $l$-th layer is defined by the consistency score between query $\mathbf{q}$ and all the neighbors as:
\begin{equation}\label{normalize}
	p^{(l)}(i;\mathbf{q})=s^{(l)}(i;\mathbf{q})/\sum_{j\in{\mathcal{N}_v}}s^{(l)}(j;\mathbf{q}).
\end{equation}
where $s^{(l)}(i;\mathbf{q})$ denotes the consistency score between the neighbor $i$ and the query $\mathbf{q}$ in $l$-th GNN layer. It is defined as: 
\begin{equation}\label{consistency score}
	s^{(l)}(i;\mathbf{q})=\text{exp}(-\|\mathbf{q}-\mathbf{h}_i^{(l)}\|_2^2),
\end{equation}
where $\mathbf{h}_i^{(l)}$ denotes the node embedding of node $i$ at $l$-th layer. For both nodes $u$ and $t$, during the inference of rating score, we use the same query embedding.
Thus, we ignore the subscript and write it as $\mathbf{q}$ for simplicity. We present this process as the sampling blocks in Figure~\ref{framework}(b), where the probabilities for neighbors are denoted as $p_i$. The number of sampled neighbors is proportional to the total number of neighbors, where the ratio is $0\leq \gamma \leq 1$. As such, we sample more neighbors if a node is connected to more nodes. 

\subsection{Relation Attention}\label{Relation Attention}


After sampling the neighbors, we should aggregate their embeddings as illustrated in Eq.~(\ref{eq:GNN}). However, the \textit{relation-level} social inconsistency suggests that we should distinguish different relations. To this end, we apply a relation attention module in \modelname for those sampled neighbors. It learns the importance of those sampled nodes by considering the associated relations. 

The relation attention assigns an importance factor $\alpha_{i}$ for each sampled node $i$. We can rewrite the AGG function in Eq.~(\ref{eq:GNN}) as:
\begin{equation}
	\text{AGG}^{(l)}=\sum_{i=1}^Q\alpha_i^{(l)} \cdot \mathbf{h}_i^{(l-1)},
\end{equation}
where $\alpha_i^{(l)}$ is the importance of the $i$-th neighbor sampled from Eq.~(\ref{normalize}) and $Q$ denotes the total number of sampled neighbors. Assuming the relation for the edge $(v,i)$ is $r_i$, we calculate $\alpha_i$ by adopting the self-attention mechanism as:
\begin{equation}
	\alpha_i^{(l)}=\frac{\text{exp}(\mathbf{w}_{a}^{\top}(\mathbf{h}_i^{(l-1)}\oplus\mathbf{e}_{r_i}))}{\sum_{j=1}^Q{\text{exp}(\mathbf{w}_{a}^{\top}(\mathbf{h}_j^{(l-1)}\oplus\mathbf{e}_{r_j}))}}
\end{equation}
where $\mathbf{e}_{r_i}\in\mathbb{R}^{d}$ represents the relation embedding of relation $r_i$ and $\mathbf{w}_{a} \in \mathbb{R}^{2d}$ is trainable parameter for the self-attention layer and $\alpha_i$ is the attention weights. We illustrate the relation attention as the green block in Figure~\ref{framework}(b). 

\subsection{Rating Prediction and Optimization}
After $L$ layer propagation, we obtain the embedding of $u$ and $t$, which are denoted as $\mathbf{h}_{u}^{(L)}$ and $\mathbf{h}_{t}^{(L)}$.
We calculate the rating score of the user-item pair $(u,t)$ by the inner-product of embeddings as:
\begin{equation}
  \hat{R}_{u,t}=\mathbf{h}_{u}^{(L)} \cdot \mathbf{h}_{t}^{(L)}.
\end{equation}

Then the loss function is defined as the Root Mean Squared Error (RMSE) between $\hat{R}_{u,t}$ and ground truth rating score $R_{u,t}$ among all $(u,t)$ pairs in $\mathcal{E}_{\text{rating}}$, which is calculated as
\begin{equation}
    \mathcal{L} = \sqrt{\frac{\sum_{(u,t)\in \mathcal{E}_{\text{rating}}}(R_{u,t}-\hat{R}_{u,t})^2}{|\mathcal{E}_{\text{rating}}|}},
\end{equation}
where $\mathcal{E}_{\text{rating}}$ is the set of all rating edges. We use Adam \cite{kingma2014adam} as the optimizer with a weight decay rate of $0.0001$ to avoid over-fitting.

\section{Experiments}
\subsection{Experimental Setup}
\subsubsection{Datasets}

Ciao and Epinions~\footnote{\url{https://www.cse.msu.edu/~tangjili/datasetcode/truststudy.htm}} are two representative datasets~\cite{tang2012etrust,tang2012mtrust,tang2013exploiting,tang2013exploiting1} for studying social recommendation problem. We remove users without social links because they are out of social recommendation scope. Ciao has $7,317$ users, $104,975$ items with $111,781$ social links. Epinions has $18,069$ users, $261,246$ items with $355,530$ social links. We also linked items that share more than $50\%$ of their neighbors.


\subsubsection{Baselines}
To justify the effectiveness of \modelname, we compare \modelname with $6$ baseline methods, including matrix factorization methods, non-GNN graph embedding methods, and GNN-based methods. SoRec \cite{ma2008sorec}, SocialMF \cite{jamali2010matrix} and SoReg \cite{ma2011recommender} incorporate social links with matrix factorization methods. CUNE~\cite{CUNE} adopts collaborative graph embedding methods.  GCMC+SN \cite{berg2017graph} and GraphRec \cite{fan2019graph} employ GNNs for learning node embeddings.

\subsubsection{Evaluation Metrics}\label{Evaluation Metrics}
To evaluate the quality of the
social recommendation, two common metrics, Mean Absolute Error
(MAE) and Root Mean Square Error (RMSE), are adopted for the rating prediction task~\cite{fan2019graph}. Note that lower values of both indicate better performance. 
And a small improvement in both may have a significant
impact on the quality of top-N recommendation~\cite{koren2008factorization}.

\subsubsection{Experimental Settings}
Each dataset is randomly split to $60\%$, $20\%$, and $20\%$ for the training, validation, and testing, respectively.
The grid search is applied for hyper-parameters tuning.
We searched neighbor percent in $\{0.2, 0.4, 0.6, 0.8, 1.0\}$.
For embedding size, we search in $\{8, 16, 32, 64, 128, 256\}$.
The learning rate is searched in $\{0.0005, 0.001, 0.005, 0.01, 0.05, 0.1\}$.
The batch size is searched in $\{32, 64, 128, 256, 512\}$.
Only one GNN layer is used for both Ciao and Epinions datasets. To cope with the over-fitting problem, early stopping was utilized in all experiments, \textit{i.e.,} stop training if the RMSE on the validation set is not improved for five epochs. 

\subsection{Performance Evaluation}

\begin{table}[htb]
\caption{Overall comparison. The best and the second-best results are in bold and underlined, respectively.}
\label{Comparison Experiment}
\begin{tabular}{lcccc}
\toprule
\multirow{2}{*}{Method} & \multicolumn{2}{c}{Ciao} & \multicolumn{2}{c}{Epinions} \\
\cmidrule(r){2-3} \cmidrule(r){4-5}
&  RMSE      &  MAE   
&  RMSE      &  MAE  \\
\midrule
SoRec  & 1.2024  & 0.8693  & 1.3389  & 1.0618  \\
SoReg  & 1.0066  & 0.7595  & 1.0751  & 0.8309  \\
SocialMF  & 1.0013  & 0.7535  & 1.0706  & 0.8264  \\
GCMC+SN & 1.0301 & 0.7970 & 1.1070 & 0.8480 \\
GraphRec  & 1.0040  & \underline{0.7591}  & 1.0799  & \underline{0.8219}  \\
CUNE  & \underline{1.0002}  & \underline{0.7591}  & \underline{1.0681} & 0.8284  \\
\modelname & \textbf{0.9722} & \textbf{0.7394} & \textbf{1.0495} & \textbf{0.8046} \\
\hline
Improvement  & 2.79\% &  1.87\% & 1.74\% & 2.1\% \\
\bottomrule
\end{tabular}
\end{table}

The experiment results of all the methods are shown in Table~\ref{Comparison Experiment}. GCMC, GraphRec, CUNE and \modelname perform better than SoRec, SoReg and SocialMF, which shows GNN and graph embedding based methods have a better capability to aggregate neighbor information.
\modelname achieves the best results on both Ciao and Epinions datasets.
It has an $1.7\%$ relative improvement on two datasets compared with the second-best one on average, which can be interpreted as a significant improvement~\cite{fan2019graph}.
The results show the benefits brought by tackling the social inconsistency problems.

\subsection{Ablation Study}
\begin{figure}
      \begin{center}
        \includegraphics[width=.22\textwidth]{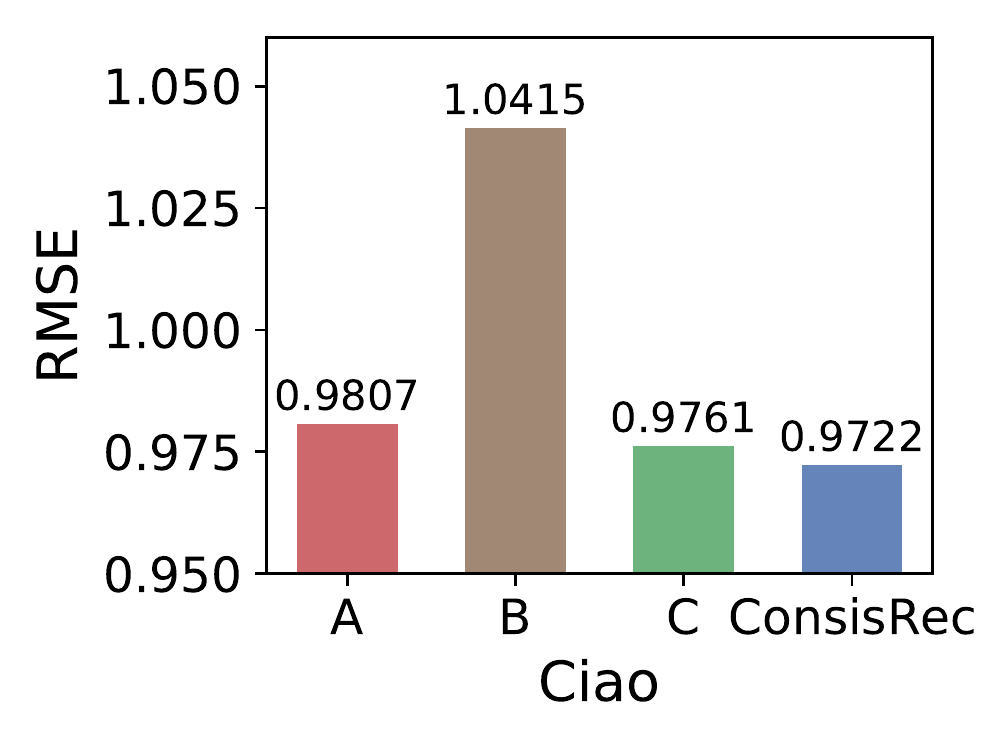}
        \includegraphics[width=.22\textwidth]{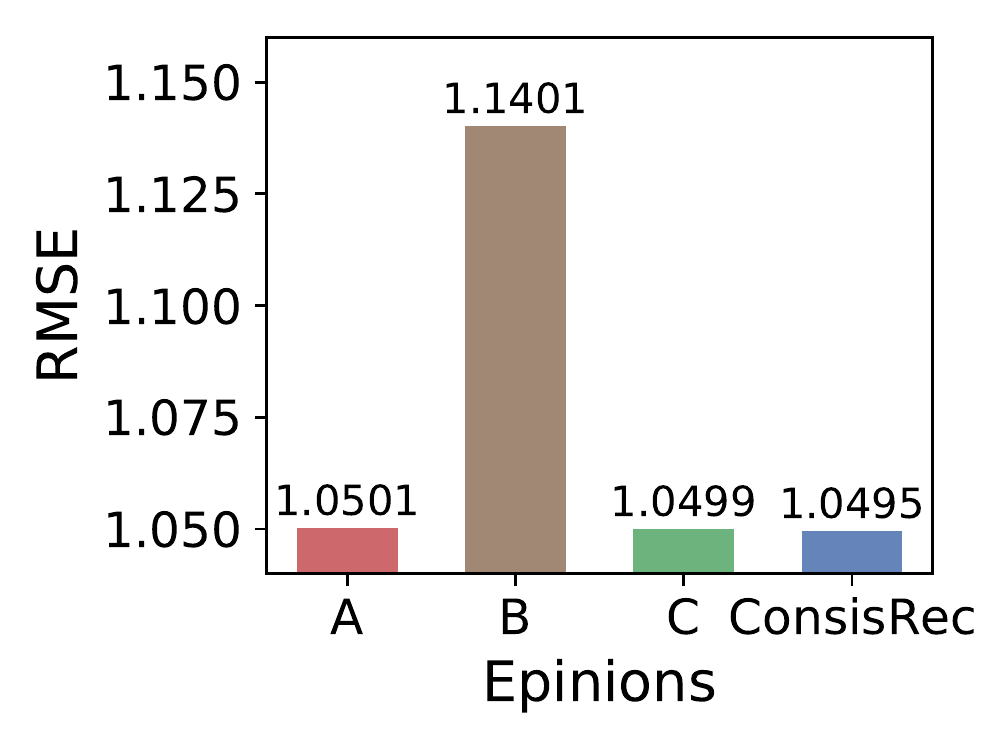}
         \includegraphics[width=.22\textwidth]{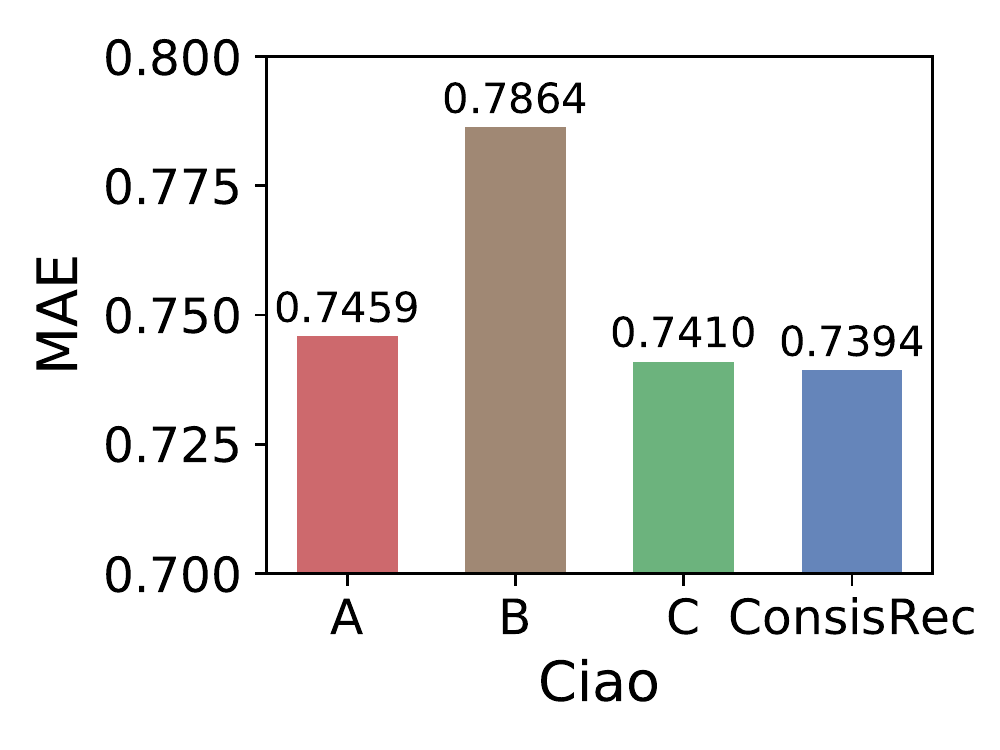}
        \includegraphics[width=.22\textwidth]{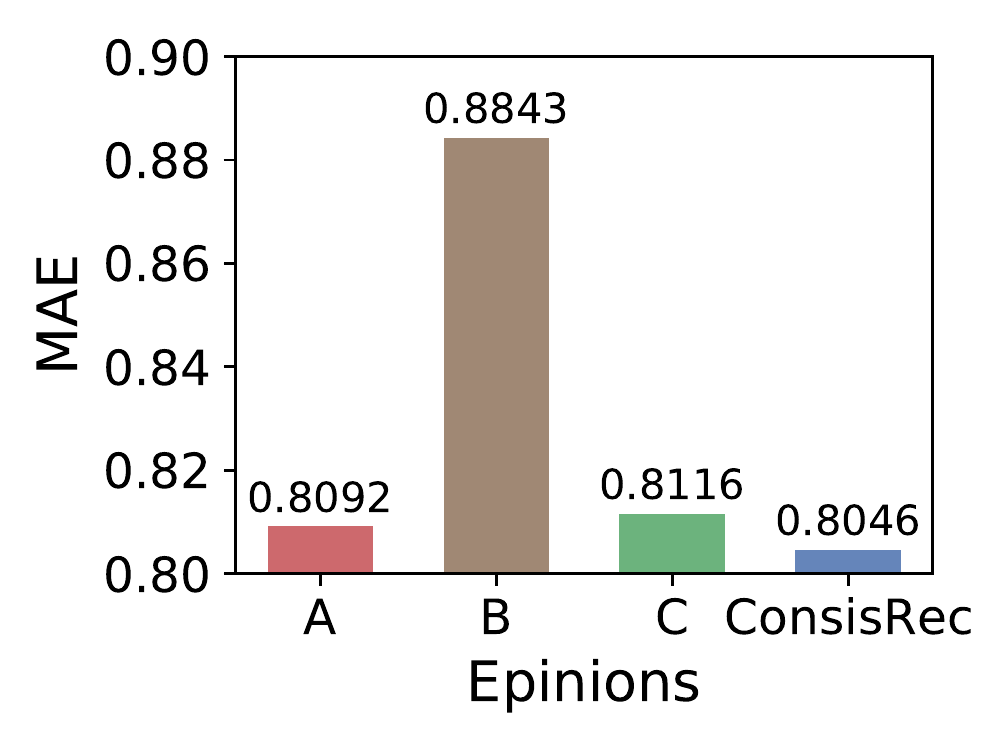}
      \end{center}
        \caption{Ablation study of \modelname.}
        \label{Ablation study}
\end{figure}

An ablation study is further made to evaluate different components in \modelname. We create three variants of \modelname, which are A, B, and C.
A is built by removing the query layer, which directly uses user embedding instead of query embedding to select the corresponding neighbors.
B is built by removing neighbor sampling, which aggregates all neighbors.
C is built by removing relation attention, which assigns equal weights to edges with different relations.
The experimental results are illustrated in Figure \ref{Ablation study}.

We can observe that \modelname consistently achieves the best performance against other variants, demonstrating that all components are necessary to yield the best results. 
Additionally, we observe that the variant $B$ (removing neighbor sampling module) dramatically spoils the performance, which justifies the importance of selecting consistent neighbors.
The worse performance of variant A and C compared with \modelname also proves the importance of query layer and relation attention, respectively.

\subsection{Parameter Sensitivity}
\begin{figure}
    \begin{center}
\includegraphics[width=.14\textwidth]{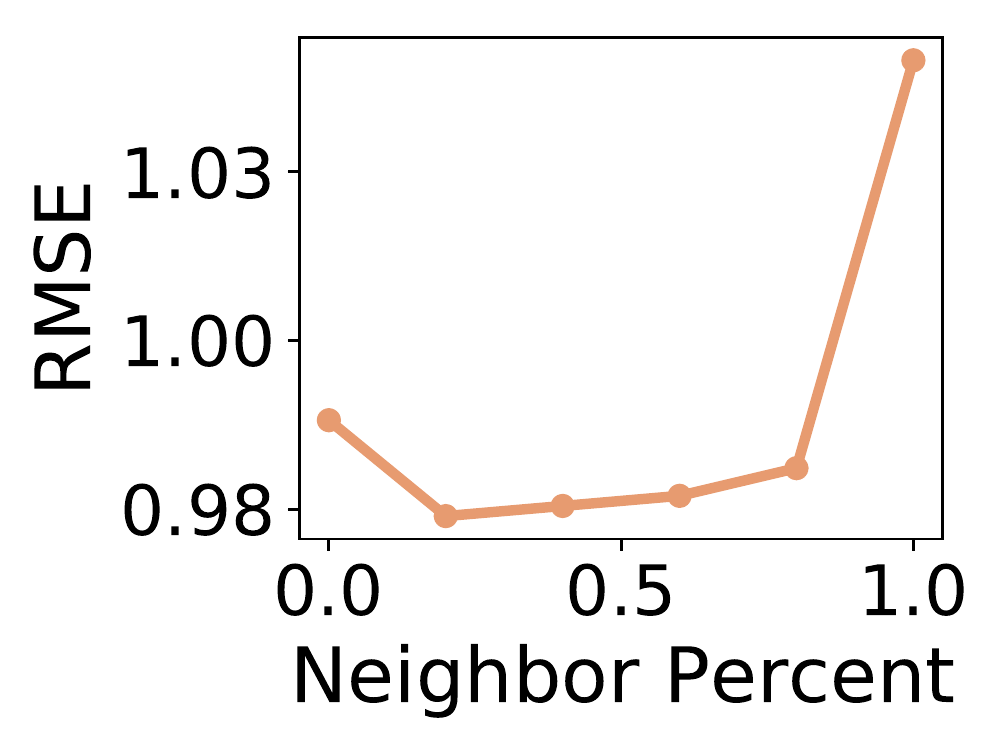}
\includegraphics[width=.14\textwidth]{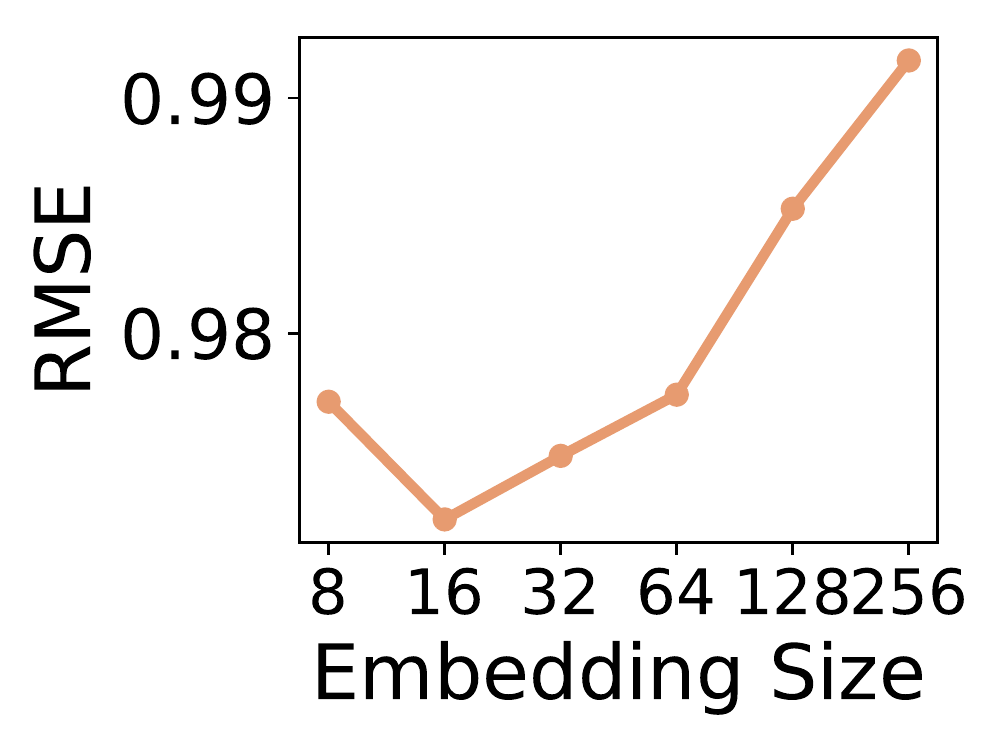}
\includegraphics[width=.14\textwidth]{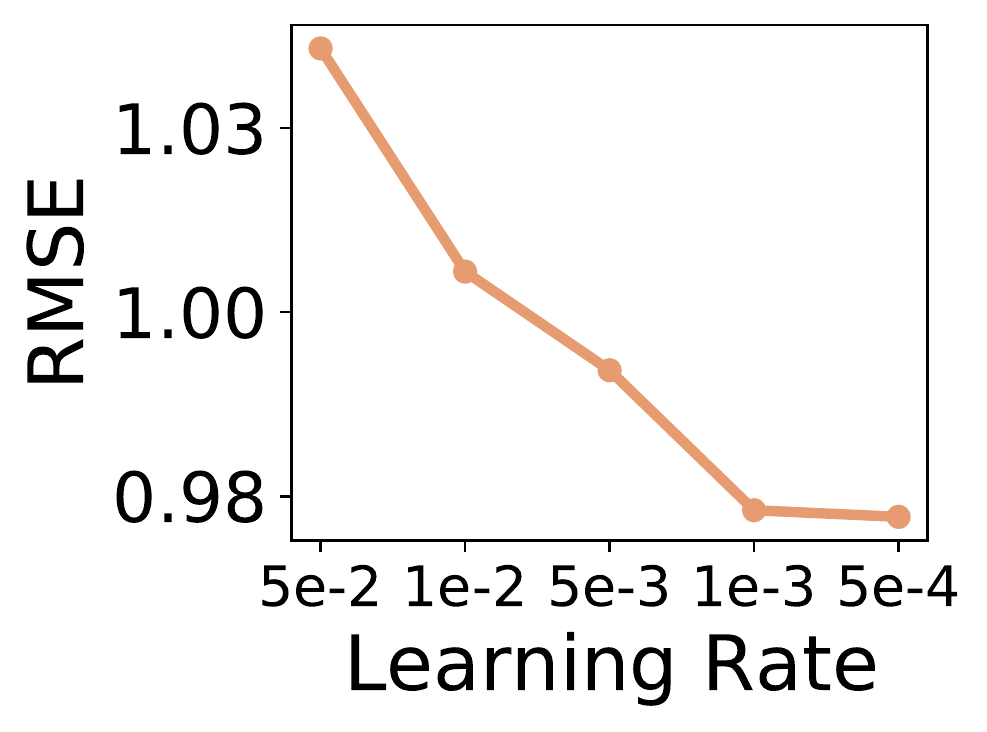}
\includegraphics[width=.14\textwidth]{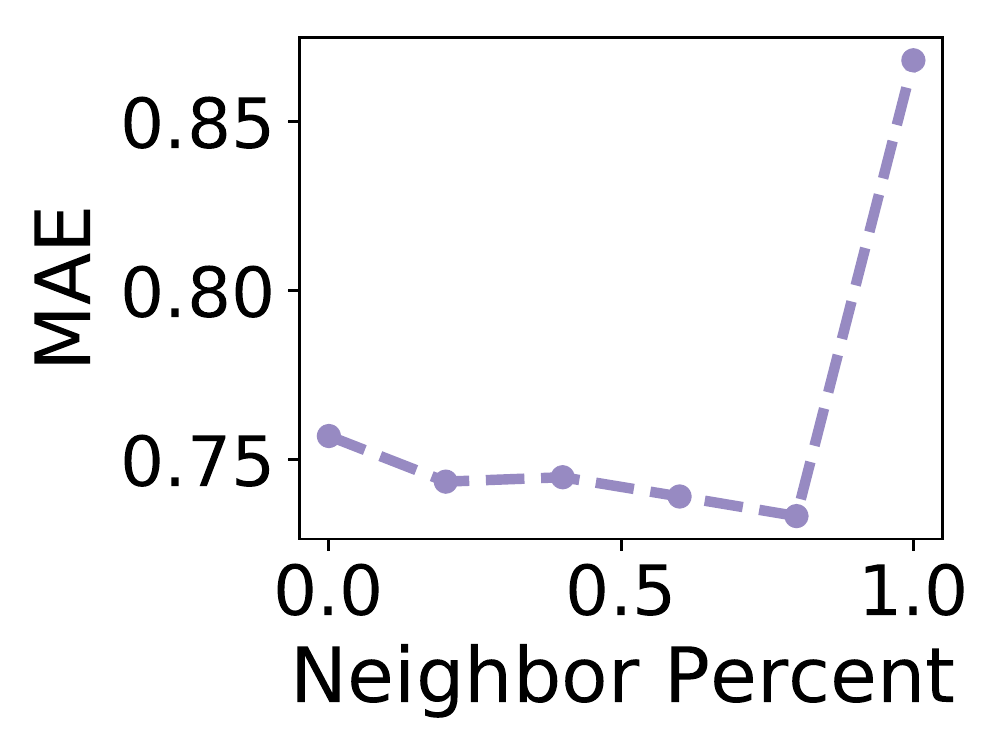}
\includegraphics[width=.14\textwidth]{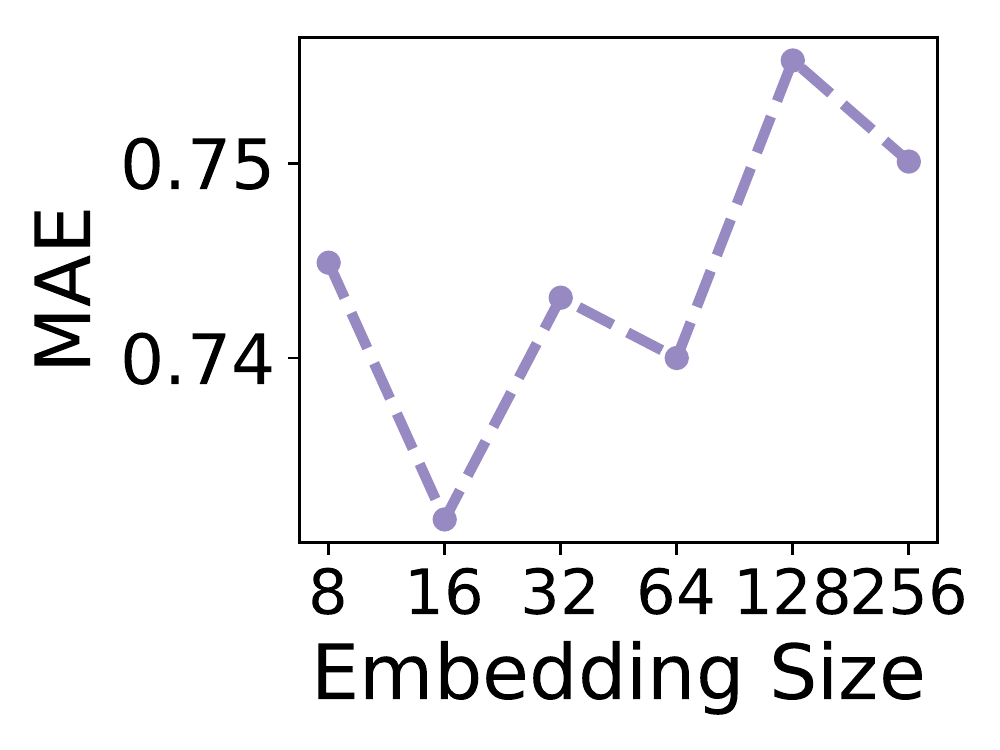}
\includegraphics[width=.14\textwidth]{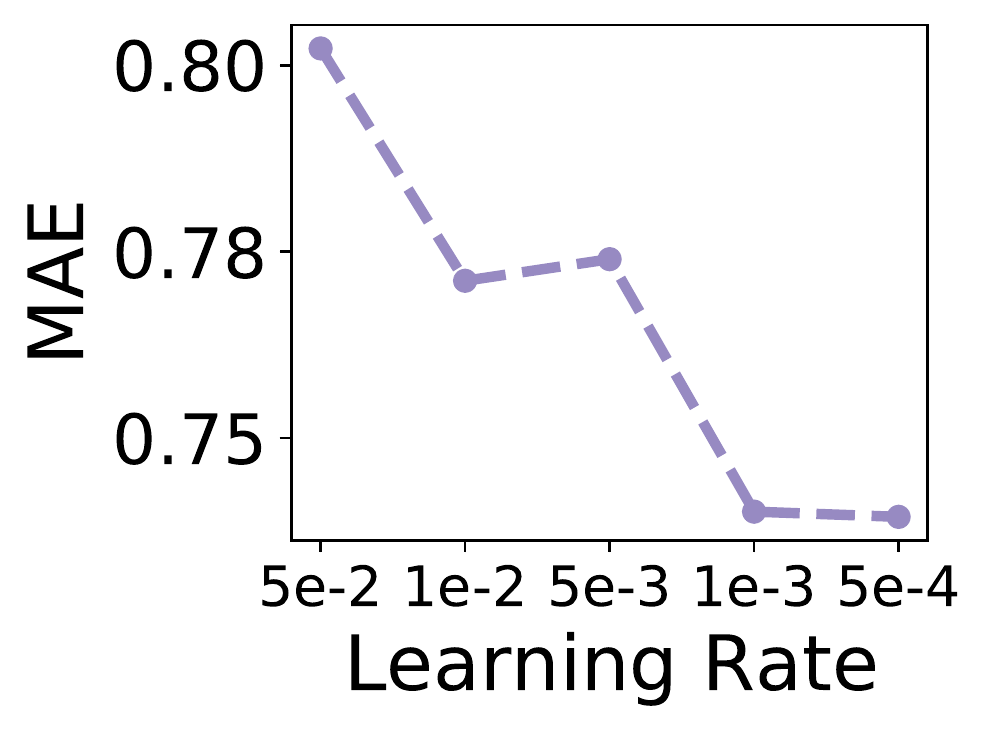}
\end{center}
    \caption{Parameter sensitivity on Ciao dataset.}
    \label{Parameter Sensitivity}
\end{figure}

Influential hyper-parameters in \modelname includes neighbor percent, embedding size and the learning rate. Due to space limitation, we only represent results on Ciao dataset in Figure \ref{Parameter Sensitivity}. For neighbor percent, we observe an obvious error increment when neighbor percent rising from $0.8$ to $1.0$, which results from aggregating those inconsistent neighbors. The best embedding size on Ciao is $16$.
Smaller embedding size is insufficient for representing node information, while large embedding size would lead to the over-fitting problem.
Learning rate has a critical impact on model performance, which needs to be tuned carefully.

\section{Conclusion and future work}
In this paper, we identify the social inconsistency problem related to social recommendation. The proposed \modelname contains three modifications on GNN to tackle the social inconsistency problem. Experiment results on two real-world datasets show the effectiveness of \modelname. Future work includes better ways to filter informative neighbors and identify the inconsistency problems inherited in other graph related research directions.

\section{Acknowledgements}
This work is supported in part by NSF under grants III-1763325, III-1909323, and SaTC-1930941. 

\newpage
\bibliographystyle{ACM-Reference-Format}
\bibliography{acmart}


\begin{thebibliography}{39}


\ifx \showCODEN    \undefined \def \showCODEN     #1{\unskip}     \fi
\ifx \showDOI      \undefined \def \showDOI       #1{#1}\fi
\ifx \showISBNx    \undefined \def \showISBNx     #1{\unskip}     \fi
\ifx \showISBNxiii \undefined \def \showISBNxiii  #1{\unskip}     \fi
\ifx \showISSN     \undefined \def \showISSN      #1{\unskip}     \fi
\ifx \showLCCN     \undefined \def \showLCCN      #1{\unskip}     \fi
\ifx \shownote     \undefined \def \shownote      #1{#1}          \fi
\ifx \showarticletitle \undefined \def \showarticletitle #1{#1}   \fi
\ifx \showURL      \undefined \def \showURL       {\relax}        \fi
\providecommand\bibfield[2]{#2}
\providecommand\bibinfo[2]{#2}
\providecommand\natexlab[1]{#1}
\providecommand\showeprint[2][]{arXiv:#2}

\bibitem[\protect\citeauthoryear{Berg, Kipf, and Welling}{Berg
  et~al\mbox{.}}{2017}]%
        {berg2017graph}
\bibfield{author}{\bibinfo{person}{Rianne van~den Berg},
  \bibinfo{person}{Thomas~N Kipf}, {and} \bibinfo{person}{Max Welling}.}
  \bibinfo{year}{2017}\natexlab{}.
\newblock \showarticletitle{Graph convolutional matrix completion}.
\newblock \bibinfo{journal}{\emph{arXiv preprint arXiv:1706.02263}}
  (\bibinfo{year}{2017}).
\newblock


\bibitem[\protect\citeauthoryear{Bizzi and Labban}{Bizzi and Labban}{2019}]%
        {bizzi2019double}
\bibfield{author}{\bibinfo{person}{Lorenzo Bizzi} {and} \bibinfo{person}{Alice
  Labban}.} \bibinfo{year}{2019}\natexlab{}.
\newblock \showarticletitle{The double-edged impact of social media on online
  trading: Opportunities, threats, and recommendations for organizations}.
\newblock \bibinfo{journal}{\emph{Business Horizons}} \bibinfo{volume}{62},
  \bibinfo{number}{4} (\bibinfo{year}{2019}), \bibinfo{pages}{509--519}.
\newblock


\bibitem[\protect\citeauthoryear{Chiang, Liu, Si, Li, Bengio, and Hsieh}{Chiang
  et~al\mbox{.}}{2019}]%
        {chiang2019cluster}
\bibfield{author}{\bibinfo{person}{Wei-Lin Chiang}, \bibinfo{person}{Xuanqing
  Liu}, \bibinfo{person}{Si Si}, \bibinfo{person}{Yang Li},
  \bibinfo{person}{Samy Bengio}, {and} \bibinfo{person}{Cho-Jui Hsieh}.}
  \bibinfo{year}{2019}\natexlab{}.
\newblock \showarticletitle{Cluster-gcn: An efficient algorithm for training
  deep and large graph convolutional networks}. In
  \bibinfo{booktitle}{\emph{Proceedings of the 25th ACM SIGKDD International
  Conference on Knowledge Discovery \& Data Mining}}.
  \bibinfo{pages}{257--266}.
\newblock


\bibitem[\protect\citeauthoryear{Cong, Forsati, Kandemir, and Mahdavi}{Cong
  et~al\mbox{.}}{2020}]%
        {cong2020minimal}
\bibfield{author}{\bibinfo{person}{Weilin Cong}, \bibinfo{person}{Rana
  Forsati}, \bibinfo{person}{Mahmut Kandemir}, {and} \bibinfo{person}{Mehrdad
  Mahdavi}.} \bibinfo{year}{2020}\natexlab{}.
\newblock \showarticletitle{Minimal variance sampling with provable guarantees
  for fast training of graph neural networks}. In
  \bibinfo{booktitle}{\emph{Proceedings of the 26th ACM SIGKDD International
  Conference on Knowledge Discovery \& Data Mining}}.
  \bibinfo{pages}{1393--1403}.
\newblock


\bibitem[\protect\citeauthoryear{Dou, Liu, Sun, Deng, Peng, and Yu}{Dou
  et~al\mbox{.}}{2020}]%
        {dou2020enhancing}
\bibfield{author}{\bibinfo{person}{Yingtong Dou}, \bibinfo{person}{Zhiwei Liu},
  \bibinfo{person}{Li Sun}, \bibinfo{person}{Yutong Deng}, \bibinfo{person}{Hao
  Peng}, {and} \bibinfo{person}{Philip~S Yu}.} \bibinfo{year}{2020}\natexlab{}.
\newblock \showarticletitle{Enhancing graph neural network-based fraud
  detectors against camouflaged fraudsters}. In
  \bibinfo{booktitle}{\emph{Proceedings of the 29th ACM International
  Conference on Information \& Knowledge Management}}.
  \bibinfo{pages}{315--324}.
\newblock


\bibitem[\protect\citeauthoryear{Fan, Ma, Li, He, Zhao, Tang, and Yin}{Fan
  et~al\mbox{.}}{2019a}]%
        {fan2019graph}
\bibfield{author}{\bibinfo{person}{Wenqi Fan}, \bibinfo{person}{Yao Ma},
  \bibinfo{person}{Qing Li}, \bibinfo{person}{Yuan He}, \bibinfo{person}{Eric
  Zhao}, \bibinfo{person}{Jiliang Tang}, {and} \bibinfo{person}{Dawei Yin}.}
  \bibinfo{year}{2019}\natexlab{a}.
\newblock \showarticletitle{Graph neural networks for social recommendation}.
  In \bibinfo{booktitle}{\emph{The World Wide Web Conference, {WWW} 2019, San
  Francisco, CA, USA, May 13-17, 2019}}. \bibinfo{publisher}{{ACM}},
  \bibinfo{pages}{417--426}.
\newblock


\bibitem[\protect\citeauthoryear{Fan, Ma, Li, Wang, Cai, Tang, and Yin}{Fan
  et~al\mbox{.}}{2020}]%
        {fan2020graph}
\bibfield{author}{\bibinfo{person}{Wenqi Fan}, \bibinfo{person}{Yao Ma},
  \bibinfo{person}{Qing Li}, \bibinfo{person}{Jianping Wang},
  \bibinfo{person}{Guoyong Cai}, \bibinfo{person}{Jiliang Tang}, {and}
  \bibinfo{person}{Dawei Yin}.} \bibinfo{year}{2020}\natexlab{}.
\newblock \showarticletitle{A Graph Neural Network Framework for Social
  Recommendations}.
\newblock \bibinfo{journal}{\emph{IEEE Transactions on Knowledge and Data
  Engineering}} (\bibinfo{year}{2020}).
\newblock


\bibitem[\protect\citeauthoryear{Fan, Ma, Yin, Wang, Tang, and Li}{Fan
  et~al\mbox{.}}{2019b}]%
        {fan2019deep}
\bibfield{author}{\bibinfo{person}{Wenqi Fan}, \bibinfo{person}{Yao Ma},
  \bibinfo{person}{Dawei Yin}, \bibinfo{person}{Jianping Wang},
  \bibinfo{person}{Jiliang Tang}, {and} \bibinfo{person}{Qing Li}.}
  \bibinfo{year}{2019}\natexlab{b}.
\newblock \showarticletitle{Deep social collaborative filtering}. In
  \bibinfo{booktitle}{\emph{Proceedings of the 13th ACM Conference on
  Recommender Systems}}. \bibinfo{pages}{305--313}.
\newblock


\bibitem[\protect\citeauthoryear{Guo and Wang}{Guo and Wang}{2021}]%
        {guo2020deep}
\bibfield{author}{\bibinfo{person}{Zhiwei Guo} {and} \bibinfo{person}{Heng
  Wang}.} \bibinfo{year}{2021}\natexlab{}.
\newblock \showarticletitle{A Deep Graph Neural Network-based Mechanism for
  Social Recommendations}.
\newblock \bibinfo{journal}{\emph{IEEE Transactions on Industrial Informatics}}
  \bibinfo{volume}{17}, \bibinfo{number}{4} (\bibinfo{year}{2021}),
  \bibinfo{pages}{2776--2783}.
\newblock


\bibitem[\protect\citeauthoryear{He, Deng, Wang, Li, Zhang, and Wang}{He
  et~al\mbox{.}}{2020}]%
        {lightgcn2020he}
\bibfield{author}{\bibinfo{person}{Xiangnan He}, \bibinfo{person}{Kuan Deng},
  \bibinfo{person}{Xiang Wang}, \bibinfo{person}{Yan Li},
  \bibinfo{person}{YongDong Zhang}, {and} \bibinfo{person}{Meng Wang}.}
  \bibinfo{year}{2020}\natexlab{}.
\newblock \showarticletitle{LightGCN: Simplifying and Powering Graph
  Convolution Network for Recommendation}. In
  \bibinfo{booktitle}{\emph{Proceedings of the 43rd International ACM SIGIR
  Conference on Research and Development in Information Retrieval}} (Virtual
  Event, China) \emph{(\bibinfo{series}{SIGIR '20})}.
  \bibinfo{publisher}{Association for Computing Machinery},
  \bibinfo{address}{New York, NY, USA}, \bibinfo{pages}{639–648}.
\newblock
\showISBNx{9781450380164}
\urldef\tempurl%
\url{https://doi.org/10.1145/3397271.3401063}
\showDOI{\tempurl}


\bibitem[\protect\citeauthoryear{Jamali and Ester}{Jamali and Ester}{2010}]%
        {jamali2010matrix}
\bibfield{author}{\bibinfo{person}{Mohsen Jamali} {and} \bibinfo{person}{Martin
  Ester}.} \bibinfo{year}{2010}\natexlab{}.
\newblock \showarticletitle{A matrix factorization technique with trust
  propagation for recommendation in social networks}. In
  \bibinfo{booktitle}{\emph{Proceedings of the 2010 {ACM} Conference on
  Recommender Systems, RecSys 2010, Barcelona, Spain, September 26-30, 2010}}.
  \bibinfo{publisher}{{ACM}}, \bibinfo{pages}{135--142}.
\newblock


\bibitem[\protect\citeauthoryear{Kingma and Ba}{Kingma and Ba}{2015}]%
        {kingma2014adam}
\bibfield{author}{\bibinfo{person}{Diederik~P Kingma} {and}
  \bibinfo{person}{Jimmy Ba}.} \bibinfo{year}{2015}\natexlab{}.
\newblock \showarticletitle{Adam: A method for stochastic optimization}. In
  \bibinfo{booktitle}{\emph{3rd International Conference on Learning
  Representations, {ICLR} 2015, San Diego, CA, USA, May 7-9, 2015, Conference
  Track Proceedings}}.
\newblock


\bibitem[\protect\citeauthoryear{Kipf and Welling}{Kipf and Welling}{2017}]%
        {kipf2016semi}
\bibfield{author}{\bibinfo{person}{Thomas~N Kipf} {and} \bibinfo{person}{Max
  Welling}.} \bibinfo{year}{2017}\natexlab{}.
\newblock \showarticletitle{Semi-supervised classification with graph
  convolutional networks}. In \bibinfo{booktitle}{\emph{5th International
  Conference on Learning Representations, {ICLR} 2017, Toulon, France, April
  24-26, 2017, Conference Track Proceedings}}.
  \bibinfo{publisher}{OpenReview.net}.
\newblock


\bibitem[\protect\citeauthoryear{Koren}{Koren}{2008}]%
        {koren2008factorization}
\bibfield{author}{\bibinfo{person}{Yehuda Koren}.}
  \bibinfo{year}{2008}\natexlab{}.
\newblock \showarticletitle{Factorization meets the neighborhood: a
  multifaceted collaborative filtering model}. In
  \bibinfo{booktitle}{\emph{Proceedings of the 14th {ACM} {SIGKDD}
  International Conference on Knowledge Discovery and Data Mining, Las Vegas,
  Nevada, USA, August 24-27, 2008}}. \bibinfo{publisher}{{ACM}},
  \bibinfo{pages}{426--434}.
\newblock


\bibitem[\protect\citeauthoryear{Li, Tei, and Fukazawa}{Li
  et~al\mbox{.}}{2020}]%
        {li2020efficient}
\bibfield{author}{\bibinfo{person}{Munan Li}, \bibinfo{person}{Kenji Tei},
  {and} \bibinfo{person}{Yoshiaki Fukazawa}.} \bibinfo{year}{2020}\natexlab{}.
\newblock \showarticletitle{An Efficient Adaptive Attention Neural Network for
  Social Recommendation}.
\newblock \bibinfo{journal}{\emph{IEEE Access}}  \bibinfo{volume}{8}
  (\bibinfo{year}{2020}), \bibinfo{pages}{63595--63606}.
\newblock


\bibitem[\protect\citeauthoryear{Lin, Sugiyama, Kan, and Chua}{Lin
  et~al\mbox{.}}{2013}]%
        {lin2013addressing}
\bibfield{author}{\bibinfo{person}{Jovian Lin}, \bibinfo{person}{Kazunari
  Sugiyama}, \bibinfo{person}{Min-Yen Kan}, {and} \bibinfo{person}{Tat-Seng
  Chua}.} \bibinfo{year}{2013}\natexlab{}.
\newblock \showarticletitle{Addressing cold-start in app recommendation: latent
  user models constructed from twitter followers}. In
  \bibinfo{booktitle}{\emph{The 36th International {ACM} {SIGIR} conference on
  research and development in Information Retrieval, {SIGIR} '13, Dublin,
  Ireland - July 28 - August 01, 2013}}. \bibinfo{publisher}{{ACM}},
  \bibinfo{pages}{283--292}.
\newblock


\bibitem[\protect\citeauthoryear{Liu, Dou, Yu, Deng, and Peng}{Liu
  et~al\mbox{.}}{2020a}]%
        {liu2020alleviating}
\bibfield{author}{\bibinfo{person}{Zhiwei Liu}, \bibinfo{person}{Yingtong Dou},
  \bibinfo{person}{Philip~S Yu}, \bibinfo{person}{Yutong Deng}, {and}
  \bibinfo{person}{Hao Peng}.} \bibinfo{year}{2020}\natexlab{a}.
\newblock \showarticletitle{Alleviating the Inconsistency Problem of Applying
  Graph Neural Network to Fraud Detection}. In
  \bibinfo{booktitle}{\emph{Proceedings of the 43rd International {ACM} {SIGIR}
  conference on research and development in Information Retrieval, {SIGIR}
  2020, Virtual Event, China, July 25-30, 2020}}. \bibinfo{publisher}{{ACM}},
  \bibinfo{pages}{1569--1572}.
\newblock


\bibitem[\protect\citeauthoryear{Liu, Fan, Wang, and Yu}{Liu
  et~al\mbox{.}}{2021}]%
        {liu2021augmenting}
\bibfield{author}{\bibinfo{person}{Zhiwei Liu}, \bibinfo{person}{Ziwei Fan},
  \bibinfo{person}{Yu Wang}, {and} \bibinfo{person}{Philip~S. Yu}.}
  \bibinfo{year}{2021}\natexlab{}.
\newblock \showarticletitle{Augmenting Sequential Recommendation with
  Pseudo-PriorItems via Reversely Pre-training Transformer}.
\newblock \bibinfo{journal}{\emph{Proceedings of the 44th international ACM
  SIGIR conference on Research and development in information retrieval}}.
\newblock


\bibitem[\protect\citeauthoryear{Liu, Li, Fan, Guo, Achan, and Yu}{Liu
  et~al\mbox{.}}{2020b}]%
        {liu2020basket}
\bibfield{author}{\bibinfo{person}{Zhiwei Liu}, \bibinfo{person}{Xiaohan Li},
  \bibinfo{person}{Ziwei Fan}, \bibinfo{person}{Stephen Guo},
  \bibinfo{person}{Kannan Achan}, {and} \bibinfo{person}{Philip~S Yu}.}
  \bibinfo{year}{2020}\natexlab{b}.
\newblock \showarticletitle{Basket Recommendation with Multi-Intent Translation
  Graph Neural Network}.
\newblock \bibinfo{journal}{\emph{arXiv preprint arXiv:2010.11419}}
  (\bibinfo{year}{2020}).
\newblock


\bibitem[\protect\citeauthoryear{Liu, Wan, Guo, Achan, and Yu}{Liu
  et~al\mbox{.}}{2020c}]%
        {liu2020basconv}
\bibfield{author}{\bibinfo{person}{Zhiwei Liu}, \bibinfo{person}{Mengting Wan},
  \bibinfo{person}{Stephen Guo}, \bibinfo{person}{Kannan Achan}, {and}
  \bibinfo{person}{Philip~S Yu}.} \bibinfo{year}{2020}\natexlab{c}.
\newblock \showarticletitle{Basconv: aggregating heterogeneous interactions for
  basket recommendation with graph convolutional neural network}. In
  \bibinfo{booktitle}{\emph{Proceedings of the 2020 SIAM International
  Conference on Data Mining}}. SIAM, \bibinfo{pages}{64--72}.
\newblock


\bibitem[\protect\citeauthoryear{Liu, Zheng, Zhang, Han, and Philip}{Liu
  et~al\mbox{.}}{2019}]%
        {liu2019jscn}
\bibfield{author}{\bibinfo{person}{Zhiwei Liu}, \bibinfo{person}{Lei Zheng},
  \bibinfo{person}{Jiawei Zhang}, \bibinfo{person}{Jiayu Han}, {and}
  \bibinfo{person}{S~Yu Philip}.} \bibinfo{year}{2019}\natexlab{}.
\newblock \showarticletitle{JSCN: Joint spectral convolutional network for
  cross domain recommendation}. In \bibinfo{booktitle}{\emph{2019 IEEE
  International Conference on Big Data (Big Data)}}. IEEE,
  \bibinfo{pages}{850--859}.
\newblock


\bibitem[\protect\citeauthoryear{Ma, Yang, Lyu, and King}{Ma
  et~al\mbox{.}}{2008}]%
        {ma2008sorec}
\bibfield{author}{\bibinfo{person}{Hao Ma}, \bibinfo{person}{Haixuan Yang},
  \bibinfo{person}{Michael~R Lyu}, {and} \bibinfo{person}{Irwin King}.}
  \bibinfo{year}{2008}\natexlab{}.
\newblock \showarticletitle{Sorec: social recommendation using probabilistic
  matrix factorization}. In \bibinfo{booktitle}{\emph{Proceedings of the 17th
  {ACM} Conference on Information and Knowledge Management, {CIKM} 2008, Napa
  Valley, California, USA, October 26-30, 2008}}. \bibinfo{publisher}{{ACM}},
  \bibinfo{pages}{931--940}.
\newblock


\bibitem[\protect\citeauthoryear{Ma, Zhou, Liu, Lyu, and King}{Ma
  et~al\mbox{.}}{2011}]%
        {ma2011recommender}
\bibfield{author}{\bibinfo{person}{Hao Ma}, \bibinfo{person}{Dengyong Zhou},
  \bibinfo{person}{Chao Liu}, \bibinfo{person}{Michael~R Lyu}, {and}
  \bibinfo{person}{Irwin King}.} \bibinfo{year}{2011}\natexlab{}.
\newblock \showarticletitle{Recommender systems with social regularization}. In
  \bibinfo{booktitle}{\emph{Proceedings of the Forth International Conference
  on Web Search and Web Data Mining, {WSDM} 2011, Hong Kong, China, February
  9-12, 2011}}. \bibinfo{publisher}{{ACM}}, \bibinfo{pages}{287--296}.
\newblock


\bibitem[\protect\citeauthoryear{Mu, Zha, He, and Tang}{Mu
  et~al\mbox{.}}{2019}]%
        {mu2019graph}
\bibfield{author}{\bibinfo{person}{Nan Mu}, \bibinfo{person}{Daren Zha},
  \bibinfo{person}{Yuanye He}, {and} \bibinfo{person}{Zhihao Tang}.}
  \bibinfo{year}{2019}\natexlab{}.
\newblock \showarticletitle{Graph Attention Networks for Neural Social
  Recommendation}. In \bibinfo{booktitle}{\emph{31st {IEEE} International
  Conference on Tools with Artificial Intelligence, {ICTAI} 2019, Portland, OR,
  USA, November 4-6, 2019}}. IEEE, \bibinfo{pages}{1320--1327}.
\newblock


\bibitem[\protect\citeauthoryear{Najork, Gollapudi, and Panigrahy}{Najork
  et~al\mbox{.}}{2009}]%
        {najork2009less}
\bibfield{author}{\bibinfo{person}{Marc Najork}, \bibinfo{person}{Sreenivas
  Gollapudi}, {and} \bibinfo{person}{Rina Panigrahy}.}
  \bibinfo{year}{2009}\natexlab{}.
\newblock \showarticletitle{Less is more: sampling the neighborhood graph makes
  salsa better and faster}. In \bibinfo{booktitle}{\emph{Proceedings of the
  Second ACM International Conference on Web Search and Data Mining}}.
  \bibinfo{pages}{242--251}.
\newblock


\bibitem[\protect\citeauthoryear{Sun, Wu, Liu, Zhu, and Chen}{Sun
  et~al\mbox{.}}{2013}]%
        {sun2013recommendations}
\bibfield{author}{\bibinfo{person}{Guang~Fu Sun}, \bibinfo{person}{Le Wu},
  \bibinfo{person}{Qi Liu}, \bibinfo{person}{Chen Zhu}, {and}
  \bibinfo{person}{En~Hong Chen}.} \bibinfo{year}{2013}\natexlab{}.
\newblock \showarticletitle{Recommendations based on collaborative filtering by
  exploiting sequential behaviors}.
\newblock \bibinfo{journal}{\emph{Journal of Software}} \bibinfo{volume}{24},
  \bibinfo{number}{11} (\bibinfo{year}{2013}), \bibinfo{pages}{2721--2733}.
\newblock


\bibitem[\protect\citeauthoryear{Sun, Wu, and Wang}{Sun et~al\mbox{.}}{2018}]%
        {sun2018attentive}
\bibfield{author}{\bibinfo{person}{Peijie Sun}, \bibinfo{person}{Le Wu}, {and}
  \bibinfo{person}{Meng Wang}.} \bibinfo{year}{2018}\natexlab{}.
\newblock \showarticletitle{Attentive recurrent social recommendation}. In
  \bibinfo{booktitle}{\emph{The 41st International {ACM} {SIGIR} Conference on
  Research {\&} Development in Information Retrieval, {SIGIR} 2018, Ann Arbor,
  MI, USA, July 08-12, 2018}}. \bibinfo{publisher}{{ACM}},
  \bibinfo{pages}{185--194}.
\newblock


\bibitem[\protect\citeauthoryear{Tang, Gao, Hu, and Liu}{Tang
  et~al\mbox{.}}{2013a}]%
        {tang2013exploiting}
\bibfield{author}{\bibinfo{person}{Jiliang Tang}, \bibinfo{person}{Huiji Gao},
  \bibinfo{person}{Xia Hu}, {and} \bibinfo{person}{Huan Liu}.}
  \bibinfo{year}{2013}\natexlab{a}.
\newblock \showarticletitle{Exploiting homophily effect for trust prediction}.
  In \bibinfo{booktitle}{\emph{Sixth {ACM} International Conference on Web
  Search and Data Mining, {WSDM} 2013, Rome, Italy, February 4-8, 2013}}.
  \bibinfo{publisher}{{ACM}}, \bibinfo{pages}{53--62}.
\newblock


\bibitem[\protect\citeauthoryear{Tang, Gao, and Liu}{Tang
  et~al\mbox{.}}{2012a}]%
        {tang2012mtrust}
\bibfield{author}{\bibinfo{person}{Jiliang Tang}, \bibinfo{person}{Huiji Gao},
  {and} \bibinfo{person}{Huan Liu}.} \bibinfo{year}{2012}\natexlab{a}.
\newblock \showarticletitle{mTrust: Discerning multi-faceted trust in a
  connected world}. In \bibinfo{booktitle}{\emph{Proceedings of the Fifth
  International Conference on Web Search and Web Data Mining, {WSDM} 2012,
  Seattle, WA, USA, February 8-12, 2012}}. \bibinfo{publisher}{{ACM}},
  \bibinfo{pages}{93--102}.
\newblock


\bibitem[\protect\citeauthoryear{Tang, Gao, Liu, and Das~Sarma}{Tang
  et~al\mbox{.}}{2012b}]%
        {tang2012etrust}
\bibfield{author}{\bibinfo{person}{Jiliang Tang}, \bibinfo{person}{Huiji Gao},
  \bibinfo{person}{Huan Liu}, {and} \bibinfo{person}{Atish Das~Sarma}.}
  \bibinfo{year}{2012}\natexlab{b}.
\newblock \showarticletitle{eTrust: Understanding trust evolution in an online
  world}. In \bibinfo{booktitle}{\emph{The 18th {ACM} {SIGKDD} International
  Conference on Knowledge Discovery and Data Mining, {KDD} '12, Beijing, China,
  August 12-16, 2012}}. \bibinfo{publisher}{{ACM}}, \bibinfo{pages}{253--261}.
\newblock


\bibitem[\protect\citeauthoryear{Tang, Hu, Gao, and Liu}{Tang
  et~al\mbox{.}}{2013b}]%
        {tang2013exploiting1}
\bibfield{author}{\bibinfo{person}{Jiliang Tang}, \bibinfo{person}{Xia Hu},
  \bibinfo{person}{Huiji Gao}, {and} \bibinfo{person}{Huan Liu}.}
  \bibinfo{year}{2013}\natexlab{b}.
\newblock \showarticletitle{Exploiting local and global social context for
  recommendation}. In \bibinfo{booktitle}{\emph{{IJCAI} 2013, Proceedings of
  the 23rd International Joint Conference on Artificial Intelligence, Beijing,
  China, August 3-9, 2013}}. \bibinfo{publisher}{{IJCAI/AAAI}},
  \bibinfo{pages}{2712--2718}.
\newblock


\bibitem[\protect\citeauthoryear{Wu, Li, Sun, Hong, Ge, and Wang}{Wu
  et~al\mbox{.}}{2020}]%
        {wu2020diffnet++}
\bibfield{author}{\bibinfo{person}{Le Wu}, \bibinfo{person}{Junwei Li},
  \bibinfo{person}{Peijie Sun}, \bibinfo{person}{Richang Hong},
  \bibinfo{person}{Yong Ge}, {and} \bibinfo{person}{Meng Wang}.}
  \bibinfo{year}{2020}\natexlab{}.
\newblock \showarticletitle{DiffNet++: A Neural Influence and Interest
  Diffusion Network for Social Recommendation}.
\newblock \bibinfo{journal}{\emph{IEEE Transactions on Knowledge and Data
  Engineering}} (\bibinfo{year}{2020}).
\newblock


\bibitem[\protect\citeauthoryear{Wu, Sun, Fu, Hong, Wang, and Wang}{Wu
  et~al\mbox{.}}{2019a}]%
        {wu2019neural}
\bibfield{author}{\bibinfo{person}{Le Wu}, \bibinfo{person}{Peijie Sun},
  \bibinfo{person}{Yanjie Fu}, \bibinfo{person}{Richang Hong},
  \bibinfo{person}{Xiting Wang}, {and} \bibinfo{person}{Meng Wang}.}
  \bibinfo{year}{2019}\natexlab{a}.
\newblock \showarticletitle{A neural influence diffusion model for social
  recommendation}. In \bibinfo{booktitle}{\emph{Proceedings of the 42nd
  International {ACM} {SIGIR} Conference on Research and Development in
  Information Retrieval, {SIGIR} 2019, Paris, France, July 21-25, 2019}}.
  \bibinfo{publisher}{{ACM}}, \bibinfo{pages}{235--244}.
\newblock


\bibitem[\protect\citeauthoryear{Wu, Sun, Hong, Fu, Wang, and Wang}{Wu
  et~al\mbox{.}}{2018}]%
        {wu2018socialgcn}
\bibfield{author}{\bibinfo{person}{Le Wu}, \bibinfo{person}{Peijie Sun},
  \bibinfo{person}{Richang Hong}, \bibinfo{person}{Yanjie Fu},
  \bibinfo{person}{Xiting Wang}, {and} \bibinfo{person}{Meng Wang}.}
  \bibinfo{year}{2018}\natexlab{}.
\newblock \showarticletitle{SocialGCN: An efficient graph convolutional network
  based model for social recommendation}.
\newblock \bibinfo{journal}{\emph{arXiv preprint arXiv:1811.02815}}
  (\bibinfo{year}{2018}).
\newblock


\bibitem[\protect\citeauthoryear{Wu, Sun, Hong, Ge, and Wang}{Wu
  et~al\mbox{.}}{2021}]%
        {wu2018collaborative}
\bibfield{author}{\bibinfo{person}{Le Wu}, \bibinfo{person}{Peijie Sun},
  \bibinfo{person}{Richang Hong}, \bibinfo{person}{Yong Ge}, {and}
  \bibinfo{person}{Meng Wang}.} \bibinfo{year}{2021}\natexlab{}.
\newblock \showarticletitle{Collaborative neural social recommendation}.
\newblock \bibinfo{journal}{\emph{IEEE Transactions on Systems, Man, and
  Cybernetics: Systems}} \bibinfo{volume}{51}, \bibinfo{number}{1}
  (\bibinfo{year}{2021}), \bibinfo{pages}{464--476}.
\newblock


\bibitem[\protect\citeauthoryear{Wu, Zhang, Gao, He, Weng, Gao, and Chen}{Wu
  et~al\mbox{.}}{2019b}]%
        {wu2019dual}
\bibfield{author}{\bibinfo{person}{Qitian Wu}, \bibinfo{person}{Hengrui Zhang},
  \bibinfo{person}{Xiaofeng Gao}, \bibinfo{person}{Peng He},
  \bibinfo{person}{Paul Weng}, \bibinfo{person}{Han Gao}, {and}
  \bibinfo{person}{Guihai Chen}.} \bibinfo{year}{2019}\natexlab{b}.
\newblock \showarticletitle{Dual graph attention networks for deep latent
  representation of multifaceted social effects in recommender systems}. In
  \bibinfo{booktitle}{\emph{The World Wide Web Conference, {WWW} 2019, San
  Francisco, CA, USA, May 13-17, 2019}}. \bibinfo{publisher}{{ACM}},
  \bibinfo{pages}{2091--2102}.
\newblock


\bibitem[\protect\citeauthoryear{Xiao, Yao, Pei, Wang, Yang, and Sheng}{Xiao
  et~al\mbox{.}}{2020}]%
        {xiao2020mgnn}
\bibfield{author}{\bibinfo{person}{Yang Xiao}, \bibinfo{person}{Lina Yao},
  \bibinfo{person}{Qingqi Pei}, \bibinfo{person}{Xianzhi Wang},
  \bibinfo{person}{Jian Yang}, {and} \bibinfo{person}{Quan~Z Sheng}.}
  \bibinfo{year}{2020}\natexlab{}.
\newblock \showarticletitle{MGNN: Mutualistic Graph Neural Network for Joint
  Friend and Item Recommendation}.
\newblock \bibinfo{journal}{\emph{IEEE Intelligent Systems}}
  \bibinfo{volume}{35}, \bibinfo{number}{5} (\bibinfo{year}{2020}),
  \bibinfo{pages}{7--17}.
\newblock


\bibitem[\protect\citeauthoryear{Zeng, Zhou, Srivastava, Kannan, and
  Prasanna}{Zeng et~al\mbox{.}}{2019}]%
        {zeng2019graphsaint}
\bibfield{author}{\bibinfo{person}{Hanqing Zeng}, \bibinfo{person}{Hongkuan
  Zhou}, \bibinfo{person}{Ajitesh Srivastava}, \bibinfo{person}{Rajgopal
  Kannan}, {and} \bibinfo{person}{Viktor Prasanna}.}
  \bibinfo{year}{2019}\natexlab{}.
\newblock \showarticletitle{Graphsaint: Graph sampling based inductive learning
  method}.
\newblock \bibinfo{journal}{\emph{arXiv preprint arXiv:1907.04931}}
  (\bibinfo{year}{2019}).
\newblock


\bibitem[\protect\citeauthoryear{Zhang, Yu, Wang, Shah, and Zhang}{Zhang
  et~al\mbox{.}}{2017}]%
        {CUNE}
\bibfield{author}{\bibinfo{person}{Chuxu Zhang}, \bibinfo{person}{Lu Yu},
  \bibinfo{person}{Yan Wang}, \bibinfo{person}{Chirag Shah}, {and}
  \bibinfo{person}{Xiangliang Zhang}.} \bibinfo{year}{2017}\natexlab{}.
\newblock \showarticletitle{Collaborative User Network Embedding for Social
  Recommender Systems}. In \bibinfo{booktitle}{\emph{Proceedings of the 2017
  {SIAM} International Conference on Data Mining, Houston, Texas, USA, April
  27-29, 2017}}. \bibinfo{publisher}{{SIAM}}, \bibinfo{pages}{381--389}.
\newblock


\end{thebibliography}

\end{document}